\documentclass{emulateapj}
\usepackage{CJKutf8}
\usepackage{graphicx,amsmath,amsthm,ulem,color}
\usepackage{natbib}

\newcommand{\be}{\begin{equation}}
\newcommand{\ee}{\end{equation}}

\newcommand{\kepler}{{\it Kepler }}
\newcommand{\icarus}{{\it Icarus }}

\def\refnew#1{(\ref{#1})}

\def\deg{^\circ}
\def\km{\rm \, km}
\def\s{\rm \, s}
\def\Kepler{{\it Kepler\,\,}}

\newcommand{\z}{\color{black}}
\newcommand{\w}{\color{black}}
\begin{document}
\begin{CJK*}{UTF8}{gbsn}

\title{Spacing of Kepler Planets: Sculpting by Dynamical Instability}
\author{Bonan Pu (濮勃南)\, \, \&\,\, Yanqin Wu (武延庆)}
\affil{Department of Astronomy \& Astrophysics, 50 St. George Street,
University of Toronto, Canada}

\begin{abstract}
  We study the orbital architecture of multi-planet systems detected by the \Kepler transit mission using N-body simulations, focussing on the orbital spacing between adjacent planets in systems showing four or more transiting planets. We find that the observed spacings are tightly clustered around $12$ mutual Hill radii, {\w when transit geometry and sensitivity limits are accounted for.}  In comparison, {\w dynamical integrations reveal} that the minimum spacing required for systems of similar masses to survive dynamical instability for as long as a billion years is, $\sim 10$ if all orbits are circular and coplanar, and $\sim 12$ if planetary orbits have eccentricities $\sim 0.02$ (a value suggested by studies of planet transit-time-variations).
  This apparent coincidence, between the observed spacing and the theoretical stability threshold, leads us to propose that typical planetary systems were formed with even tighter spacing, but most, except for the widest ones, have undergone dynamical instability, and are pared down to a more anemic version of their former selves, with fewer planets and larger spacings.  So while the high multiple systems (five or more transiting planets) are primordial systems that remain stable, the single or double planetary systems, abundantly discovered by the \Kepler mission, may be the descendants of more closely packed high multiple systems.  If this hypothesis is correct, we infer that the formation environment of \Kepler systems should be {\w more dissipative than that of the terrestrial planets.}
 \end{abstract}

\section{Introduction} 
\label{sec:intro}

NASA's \kepler mission has discovered that many of the planetary systems around solar-type stars contain multiple planets \citep[Kepler multis,][]{Borucki,2011Natur.470...53L,2013ApJS..204...24B}. Planets detected in these systems likely possess low mutual inclinations {\w \citep[e.g.][]{Lissauer2011b,TremaineDong,FangMargotInclination,Fabrycky14}}, and are spaced closely.  In multiple systems that contain high number of planets (4-6), the spacing is so tight that there may be of order $\sim 10$ planets fitted inside the orbit of Mercury.

Systems where a single planet transit (Kepler single) and where multiple planets transit (Kepler multis) may share the same intrinsic architecture but only differ in transit geometry.  However, a few statistical studies have shown that this is not the case.  There appears to be an excess in the number of singles over what one expects by extrapolating from the Kepler multis \citep{2011Natur.470...53L,Johansen,Weissbein,Ballard}, suggesting that many of the singles are intrinsically singles. Even the multis may contain different sorts of systems. 
A direct probe of the underlying population is provided by transit-time-variations (TTV).  \citet{XWL} probed the presence of close-companions (transiting or not) around Kepler candidates using TTV signals. They found a striking correlation between the fraction of planets that show detectable TTVs and their transit multiplicity. Systems where four or more planets transit enjoy four times higher TTV fraction than Kepler singles do, and about twice as high as those with two or three transiting planets.  This led them to propose that there are at least two different classes of Kepler systems, one closely packed and one sparsely populated \citep[also see][]{Weissbein}. Besides from planet spacing, orbital eccentricity and inclinations also differ between the two classes.  Compared to Kepler multis, the transit durations of Kepler singles appear to be longer, requiring their orbits to have higher eccentricities {\w (Wu, unpublished)}; study of stellar obliquity suggests that Kepler singles may lie on orbits more misaligned with the stellar spin than the multis \citep{MortonWinn}.  Both lend to the picture that Kepler multis are dynamically cold systems, while singles are dynamically hotter.

What could have produced such a dichotomy? \citet{Johansen} explored the possibility of dynamical instability. They boosted the masses of observed triple planetary systems by about a factor of 10 and found that these systems could become destabilized within 10 Billion years. This led them to hypothesize that Kepler singles may be the aftermath of dynamical instability. In this study, we provide much more definite arguments to this hypothesis by investigating the spacing distribution among the closely packed multi systems (4 planets and more).  We find that, adopting realistic planetary masses and ages, Kepler high multiple systems are packed close to the boundary of stability.\footnote{In contrast, the lower multiple systems are much more loosely packed \citep[including the triple systems studied by][]{Johansen}.} This proximity has to be explained.  Unless the formation mechanism is somehow capable of producing this proximity, the most likely explanation is that the primordial systems had a range of spacing, but the ones that were packed too closely have been sculpted away by the subsequent billions of years of evolution.  In this hypothesis, Kepler singles and lower multiples are the descendants of Kepler multis. If this true, we may be able to recover the primordial spacing of planetary systems. 

\section{How tight can you pack planets? -- theoretical expectations}
\label{sec:tight}

\subsection{Previous Works}
\label{subsec:previous}

Here, we set the scene by reviewing some previous results on stability in multiple planetary systems.

Johannes Kepler (1571-1630), after whom the \kepler mission is named,
initially believed that the spacing of planets in the Solar system is limited by packing perfect solids (polygons with identical faces). He discarded this idea after developing his famous Kepler laws. However, the question still remains, how tight can you pack planets together without affecting dynamical instability. Over the past few decades, some theoretical progress have been made in answering this question, and in some special cases the answers are known exactly.

One usually measure spacing between every two planets in a unit called
the mutual Hill sphere
\begin{equation}
R_{H} = \frac {a_1 + a_2}{2}\left(\frac{m_1+m_2}{3M_*}\right)^{1/3},
\label{eq:RH}
\end{equation}
where $a_i$, $m_i$ are the semi-major axis and mass of the two planets, and $M_*$ the mass of the central star.  In this unit, planet spacing can be expressed as 
a dimensionless number $K$
\begin{equation}
K = {{a_2 - a_1}\over{R_H}} .
\label{eq:K}
\end{equation}

We first focus only on orbits that are circular and coplanar.  A well known result, derived by \citet{Wisdom} based on overlapping of first order mean-motion resonances, is that a test particle ($m_2 = 0$) perturbed by a planet ($m_1, a_1$) is subject to orbital chaos if it lies too close to the planet,
\begin{equation}
{{|a_2 - a_1|}\over{(a_2+a_1)/2}} \leq 1.49 \mu^{2/7}\, .
\label{eq:wisdom}
\end{equation}
where {\w $\mu \equiv \frac{m_1+m_2}{M_*}$} is the planet mass ratio and the numerical coefficient is adopted from \citet{Duncan}. In $K$ value, the critical spacing is \begin{equation}
K_{\rm crit} = 3.46 \left({{\mu}\over{4.5\times 10^{-5}}}\right)^{-({{1}\over{21}})}\, .
\label{eq:wisdomK}
\end{equation}
This result is little modified if the test particle is instead substituted by a planet \citep{Deck}.  The dependence of $K_{\rm crit}$ on $\mu$ (albeit weak) indicates that Hill sphere is not necessarily the best unit to measure stability. However, we adopt it here for convenience.
There is also an important result on Hill stability (stability against orbit crossing), 
popularized in astronomy by \citet{Gladman}. 
As a result of total energy and angular momentum conservation, two initially circular, coplanar planets can avoid close encounters at all times if they are separated by more than $K_{\rm crit} = 3.46$. These two criteria meet at $\mu = 4.5\times 10^{-5}$ and their physical connection is discussed in \citet{Deck}.
 
For systems where there are more than 2 planets, there is yet no theoretical understanding for the origin of dynamical instability \citep[see attempts by, e.g.][]{Chambers,Nanda,Quillen}. Whatever information we have on how tight one can pack planets in these cases are empirical results from numerical simulations {\w \citep{Chambers,YKM,Nanda,Chatterjee08,SmithLissauer,Funk}}. These we summarize below.

Generally speaking, systems with more planets ($N$) require larger $K$ spacing to stay stable.  
To compare the various numerical studies, we scale the physical timescale by the orbital period of the inner most planet in the system, $\tau = t/T_1$ \citep[also see][]{Funk}. We also suppress error-bars in the fitting results but caution that there are intrinsically large scatters.

\begin{itemize} 

\item dependence on $\tau$:
Two-planet systems can remain indefinitely stable as long as they are above a critical spacing. This may not hold for large-N systems. It appears that a larger spacing does not guarantee eternal peace but only delays the time of instability (we return to this point in \S \ref{subsec:extrapolate}). Numerically, the critical spacing at which an average system undergoes their first encounter at time $\tau$ can be modelled as
\citep{Chambers,SmithLissauer,Funk},
\begin{equation}
K_{\rm crit} = a + b\log \tau \, ,
\label{eq:kcrit}
\end{equation}
 though the alternative scalings $\log K = a + b \log \tau$ and {\z$K = c \log[b (\log \tau - a)] $ have been adopted by \cite{Nanda} and \cite{Chatterjee08} respectively}. The numerical coefficients $a$ and $b$ depend on system parameters like the number of planets, their masses, and orbital eccentricities and inclinations. Two results most relevant to our case are
\begin{equation}
 K_{\rm crit} \approx (1.33\pm 0.3) \log \tau + (0.0\pm 2.0) \, .
\label{eq:tauK}
\end{equation}
for systems of $8$ planets all with $\mu=5\times 10^{-5}$ 
\citep{Funk},
and
\begin{equation}
K_{\rm crit} \sim \log\tau + 1.7 \,
\label{eq:tauK2}
\end{equation}
for $5$ planet systems with $\mu= 3\times 10^{-6}$ \citep{SmithLissauer,Nanda}.
Both scalings are obtained for circular, coplanar systems with a uniform planetary mass and a uniform K-spacing.

\item dependence on $N$: 
The most comprehensive study to date on this issue is that by \citet{Funk}. 
They found,  for $\tau = 10^{8.5}$, 
and $\mu = 5\times 10^{-5}$,
the critical separation for a system of circular, coplanar planets
goes as
\begin{equation}
K_{\rm crit} \approx \begin{cases}  4+(N-2) = 2+N &\mbox{if } N < 8 \\
10 & \mbox{if } N \ge 8 \,. 
\end{cases}
\label{eq:Ndep}
\end{equation}
This results roughly recovers equation \refnew{eq:wisdomK} when $N=2$. The initial rise of $K_{\rm crit}$ with $N$ is intuitively understood as the presence of more planetary perturbers exerts more strain on the delicate dynamical systems. The subsequent flattening of $K_{\rm crit}$ with $N$ is, however, surprising and not understood. One guess is that dynamical instability is most affected by close neighbours, and the presence of far-away neighbours, once beyond a certain separation, do not matter.  For instance, a system of 8 Neptunes will space a period range of $\approx 10$.
With such a large spacing, newly added planets have no chance of participating in any MMRs with the pre-existing planets, and therefore should not adversely affect the stability of the system.

\item dependence on $\mu$: It is found that less massive planets, with their smaller Hill spheres, need be separated by slightly higher values of $K$. Both \citet{Chambers} and \citet{Nanda} found that, for every decade of drop in $\mu$ (within the range of $\mu=10^{-9}$ to $\mu=10^{-3}$), 
the critical separation $K$ should increase by $0.5-1$.

\item Eccentricity/Inclination: If the planetary orbits are dynamically hot, more elbow room is needed to maintain stability \citep[e.g.][]{YKM}.
\citet{Nanda} showed that, at low eccentricities, increasing the average eccentricity by $0.01$ raises $K_{crit}$ by $\sim 1$. We know of no study that explores the impact of mutual inclination separately, so we provide some illustrations of this below.

\end{itemize}


Returning to our problem at hand -- a system of 8 flat, circular planets, with masses of $\sim 5 M_\oplus$, and an inner orbit of $0.1$ AU, may remain stable for over 1 Gyrs 
(or $\tau \geq 10^{10.5}$), if $K \sim 14 \pm 5$ (extrapolating from equation \ref{eq:tauK}).
This valueis broadly consistent with the spacing observed in tight Kepler systems. However, a more refined approach is required.


\subsection{Our numerical explorations}

To compare theoretical results to Kepler systems, we need to overcome a few restrictions of previous studies.  First, in almost all previous numerical simulations, planets in a given systems are assigned identical masses and identical spacing, in contrast to realistic systems.  Inhomogeneities in mass and spacing may impact results significantly, e.g., they will introduce large scatter in stability lifetimes, among systems that have comparable bulk parameters.  It is therefore necessary to study system stability in a statistical sense -- {\w previous investigations have studied only a small number of cases (typically fewer than a hundred systems).} We accomplish this by integrating a much larger number of systems ($\sim 10^4$).
In addition, we wish to study the impacts of small eccentricity and inclination. Lastly, our simulations run for a factor of $10$ longer than previous works, up to $10^8$ years.  Although still shorter than the typical system ages (a Giga-year), they provide a better indicator for long-term stability.


\subsubsection{Numerical Setup}

We generate artificial planetary systems as follows: each system contains $7$ planets orbiting around a 1 solar mass star, on coplanar and circular orbits. The number $7$ is chosen for a number of reasons: 
this is the largest number of transiting planets in a multiple system {\w \citep[e.g.][]{Lissauer2014}} ; it is close enough to the plateau in equation \refnew{eq:Ndep}, the results may be considered valid for all systems with more than $7$ planets; it is computationally feasible.  The mass of the host star is taken to be {\w $M_* = M_\odot$}, while the mass of the planets are drawn from a uniform distribution between $3.0M_E \ge m \ge 9.0M_E$, consistent with the range of masses observed in Kepler systems \citep{WuLithwick,HaddenLithwick,Weiss}.  {\w The latter assumption is robust as we know the stability threshold only depends weakly on the planet masses (\S \ref{subsec:previous}).}
The inner most planet is placed at $0.1$ AU. {\w We then determine the spacing, in unit of mutual Hill sphere, for an} adjacent pair of planets {\w by drawing} from a normal distribution with mean $K_{\rm mean}$ and variance $\sigma_K$, \begin{equation} P(K) dK = {1\over{\sigma_K \sqrt{2 \pi}}} e^{-{{(K-K_{\rm mean})^2}\over{2 \sigma_K^2}}}\, dK\, . \label{eq:probK} \end{equation}
The initial arguments of periapsis and the mean motions are drawn from a uniform distribution in $[0,2\pi]$ . 

 
We adopt the Wisdom-Holman integration scheme \citep{1991AJ....102.1528W}, as implemented in the SWIFTER package by David E. Kaufmann, an improved version of the original SWIFT package by Levison and Duncan. The post-Newtonian correction is added as an additional potential of the form $U_{GR} = -3(G M_{\star}/c r)^2$, and we adopt a time-step of $10^{-3}$ years. Integrations are discontinued
either when two planets approached one another within one mutual Hill radius (and therefore are destined for a close encounter), or when a preset time limit (usually $10^7$ years) is reached. 
We record the stability time of each system, and analyze how it depends on various parameters.
Of particular interest is the survival probability at a given time.

\subsubsection{Results for Flat, Circular Systems}

{\w We integrate, up to a time of $10^7$ years, a total of 8,000 circular, coplanar systems ($\sigma_{e} = \sigma_i = 0$) with $6.0 \le K_{mean} \le 13.0$ and $0 \le \sigma_{K} \le 4.0$. In addition, we simulate 120 such systems, within the range $6.0 \le K_{mean} \le 13.0$ and $3.25 \le \sigma_{K} \le 3.75$, up to a time of $10^8$ years. }

\begin{figure}
\includegraphics[width=0.49\textwidth,trim=0 0 0 0, clip]{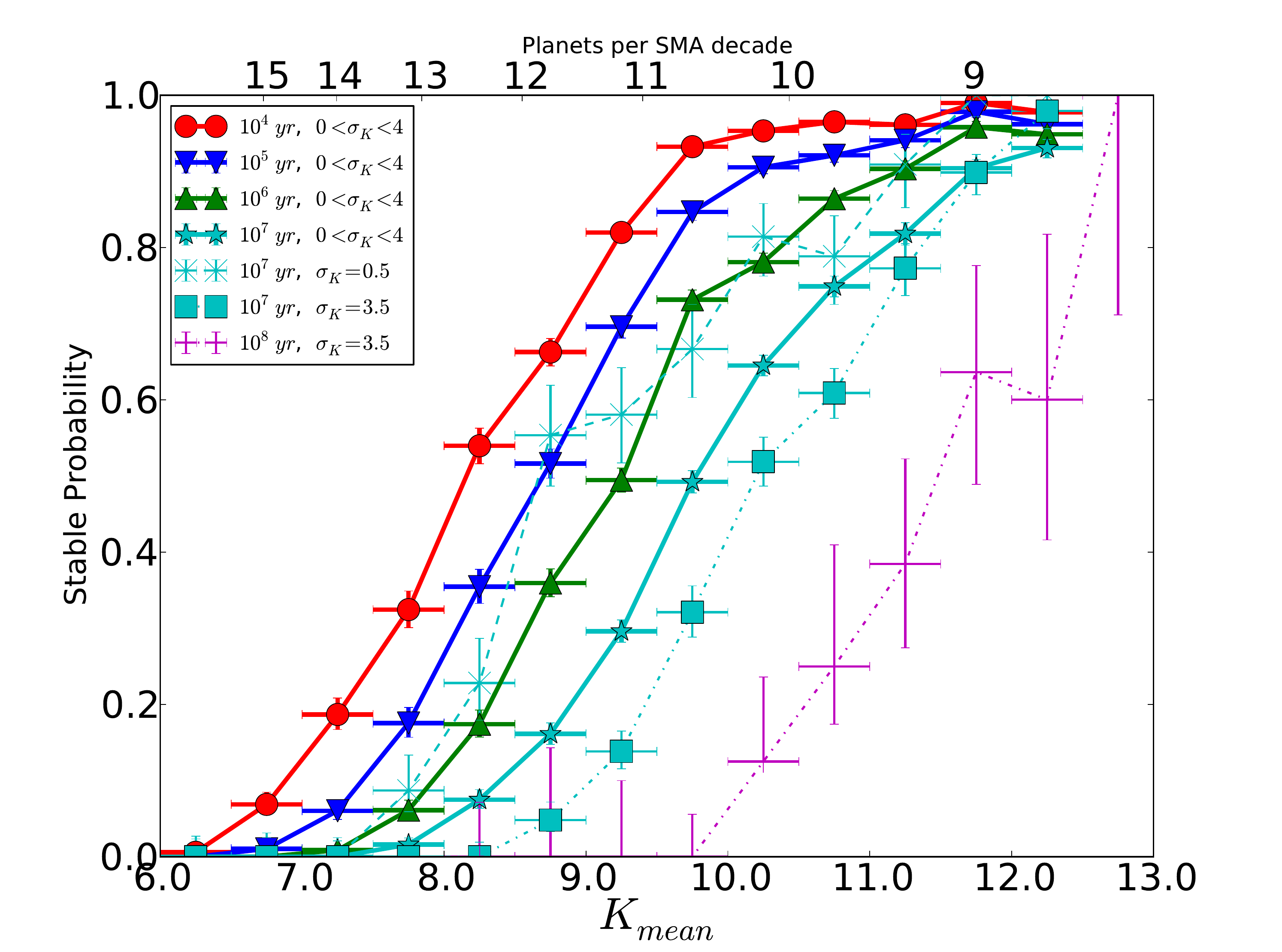}
\caption{Survival probability, measured at time $t$ (with $t$ ranging from $10^4$ to $10^8$ yrs), for systems with different $K_{\rm mean}$.  Each point represents results from some $500$ systems, with its horizontal error bars representing the width in $K_{\rm mean}$ 
  and vertical error bars accounting for Poisson error in numbers. The solid lines show the aggregated results for a distribution of $\sigma_K$. The latter is taken to be a Gaussian distribution with a mean of $2$ and a dispersion of $\sim 1$,
  with the exception of the black curve where $3.25 \le \sigma_{K} \le 3.75$. It appears that $K_{\rm mean}$ is a reasonable indicator for system stability. However, system stability also depends on the dispersion in $K$. As the two curves (dashed and dotted, both for $t=10^7$ yrs but each with a different $\sigma_K$) show, systems with larger dispersion tend to be more unstable.  So simply measuring the mean does not reveal the whole picture.}
\label{fig:figure3}
\end{figure}

We report the survival probability, the fraction of systems that remain stable up to a time $t$, as a function of planet spacing. Overall, a wider $K_{\rm mean}$ leads to a higher survival probability (Fig. \ref{fig:figure3}), however, the dispersion in $K$ ($\sigma_K$) also impacts stability.  For the same $K_{mean}$, systems with a greater $K$ dispersion amongst its pairs are less stable. In other words, a pair of planets with a smaller-than-average $K$ have a destabilizing effect on the whole system that out-weighs any corresponding stabilizing effect afforded by a planet pair with a larger-than-average $K$.

We find, empirically, that our results can be succinctly summarized using an effective spacing, \begin{equation} K_{\rm eff} \equiv K_{\rm mean} - 0.5 \sigma_K\, .  \label{eq:keff} \end{equation}

We arrive at this combination by noticing that, in our simulations, increasing $\sigma_{K}$ by 1.0 has the same effect on stability as decreasing $K_{mean}$ by $0.5$. So for systems with a varied spacing, $K_{\rm eff}$ acts as an equivalence of $K$ in systems with a uniform spacing. Such a combination can also explain the results in \citet{Chambers} where they made some preliminary explorations on the impact of a $K$-dispersion.

\begin{figure} \includegraphics[width=0.49\textwidth,trim=0 0 0 0,clip]{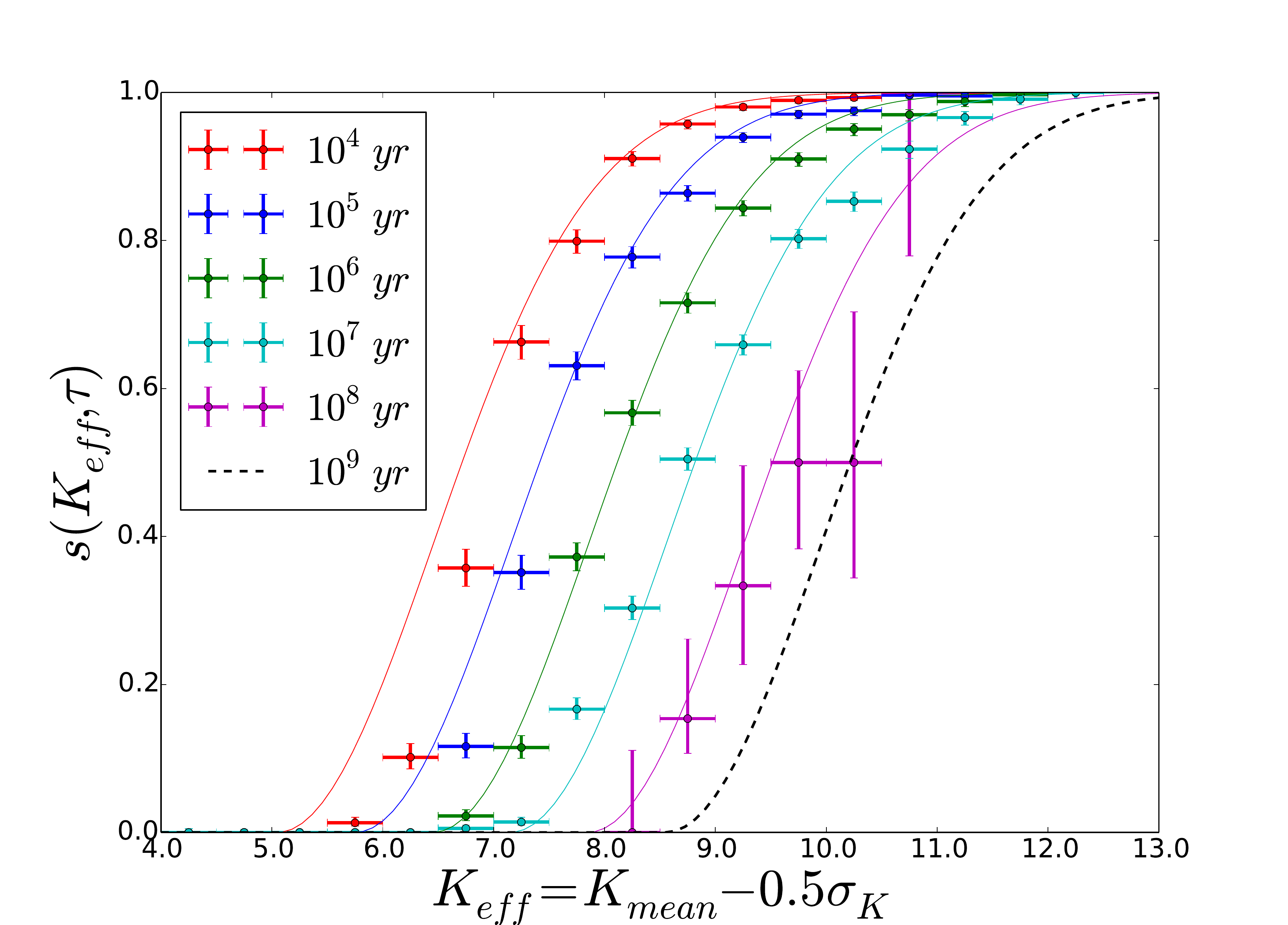} \caption{ Same as Fig. \ref{fig:figure3} but with the results plotted against $K_{\rm eff} = K_{\rm mean} - 0.5 \sigma_K$. This combination of parameters is found to describe the numerical results succinctly. Different solid lines represent equation \refnew{eq:sk}, evaluated at their respective times.  Dashed curve shows the extrapolation of equation \refnew{eq:sk} to $t = 10^9$ yrs (or $\tau = 10^{10.5}$). 
} \label{fig:figure4} \end{figure}

We can now fit the survival probability  reported in Fig. \ref{fig:figure4} as a function of 
$K_{\rm eff}$ (the functional form being somewhat arbitrary)
\begin{equation}
{\w s(K_{\rm eff},\tau) = 1 - e^{-\left[{0.5{K_{\rm eff}}} - 0.6 - 0.35\log \tau\right]^{2}}\, ,}
\label{eq:sk}
\end{equation}
for $K_{\rm eff} \geq 1.2 + 0.7 \log \tau$ and $0$ otherwise. 
Here the dimensionless time $\tau = t/T_1$, and $T_1$ is the orbital period of the inner most planet ($\sim 12$ days in our case).
%
We now define a threshold spacing, $K_{\rm 50}(t)$, as the smallest $K_{\rm eff}$ necessary to achieve a survival probability of $50\%$ at time $t$. This spacing has a similar functional form to that presented in \citet{Chambers,SmithLissauer,Funk}, but with different numerical coefficients, \begin{equation} 
K_{\rm 50}(t) \approx 0.7 \log \tau + 2.87\, .  
\label{eq:K50} \end{equation}

\begin{figure}
\includegraphics[width=0.45\textwidth,trim=0 110 50 250, clip]{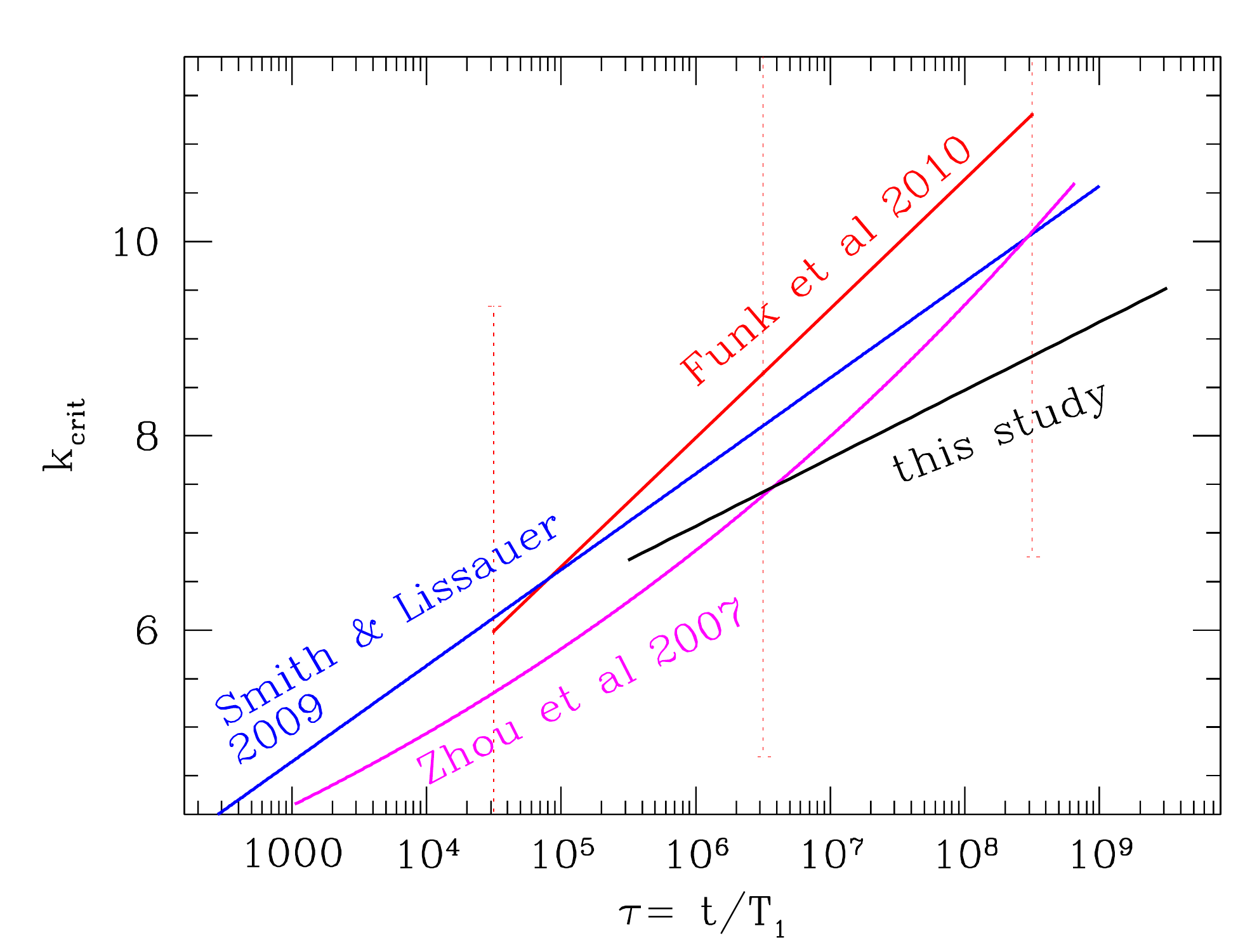}
\caption{The critical $K$ spacing required to maintain stability over time $\tau$ (x-axis, in logarithm), as obtained by different groups.  The \citet{Funk} results (in red, equation \ref{eq:tauK}, 
with vertical error bars) are for $\mu=5\times 10^{-5}$; the \citet{SmithLissauer} results (in blue, equation \ref{eq:tauK2}) applies to $\mu=3\times 10^{-6}$; we insert $\mu=2\times 10^{-5}$ into the fitting formula of \citet{Nanda}, and modify their expression to suit present notation, to
 obtain the magenta line. All these simulations are of uniform mass and uniform spacing {\w and therefore are not statistical in nature}.  Our results, $K_{\rm 50}$ (eq. \ref{eq:K50}), are plotted as a black curve. The horizontal extent of each curve also delineates the respective time range probed by different studies -- the present study probes up to$\tau = 10^{8.5}$,  ten times longer than previous ones. 
Our critical spacing are broadly consistent, though somewhat smaller, 
than those of previous studies. }
\label{fig:kcollection}
\end{figure}

We compare our critical spacing with those from previous works in Fig. \ref{fig:kcollection}. {\w
Our results, with a spread in $K$ and in planetary masses, have to be characterized by $K_{\rm 50}$. This differs from the definition of $K_{\rm crit}$ for other groups, which typically adopt a uniform $K$ and mass. However,  results} broadly agree within error-bars \citep[as given by][]{Funk} but ours tend to fall below. In other words, at the same spacing (measured by $K_{\rm eff}$ in our cases), inhomogeneous systems (with mass and spacing dispersions) remain stable slightly longer than homogeneous systems do. We note that homogenous systems, with their uniform period ratios, tend to be strongly affected by the presence of mean-motion resonances \citep[see, e.g.]{SmithLissauer,Funk}. This may partially explain the difference. A more important difference between homogenous and inhomogeneous systems is that the former may reach eternal stability at a finite spacing \citep[as suggested by][]{SmithLissauer}, while we do not observe this for the latter (see \S \ref{subsec:extrapolate}). 
From now on, we consider that our results (equation \ref{eq:K50}) supercedes equations \refnew{eq:tauK}-\refnew{eq:tauK2}.

Our scaling is valid for 4 orders of magnitude in $\tau$ ($\tau = 10^{5.5}$ to $10^{9.5}$, Fig. \ref{fig:figure4}). In subsequent sessions, we assume that it remains valid for another one order of magnitude, to $\tau = 10^{10.5}$ (or $t =$ 1 Gyrs). At this epoch, the critical spacing is 
$K_{\rm 50} \sim 10.2$.


\subsubsection{Results for Eccentric, Inclined Systems}

Here, we explore quantitatively how eccentricity and inclination affect the critical spacing.
We generate 12,000 artificial 7-planet systems with a normal distribution in spacing, $6.0 \le K_{mean} \le 20.0$, $3.0 \le \sigma_{K} \le 4.0$, and a Rayleigh distribution in eccentricity and inclination. The scale factors are $\sigma_e$ and $\sigma_{\rm inc}$, e.g.,
\begin{equation} P(e) de = {{e}\over{\sigma_e^2}} e^{-{{e^2}\over{2\sigma_e^2}}} \, de\, .  \label{eq:probe} \end{equation}
We explore the range $\sigma_e \in [0, 0.05]$, 
and $\sigma_{\rm inc} \in [0, 0.16]$ (in radian, corresponding to a range of $[0^o,9^o]$). 
The initial longitudes of ascending nodes are randomly distributed between $0$ and $2\pi$.

\begin{figure}
\centering
\includegraphics[width=0.49\textwidth,trim=0 0 0 0,clip]{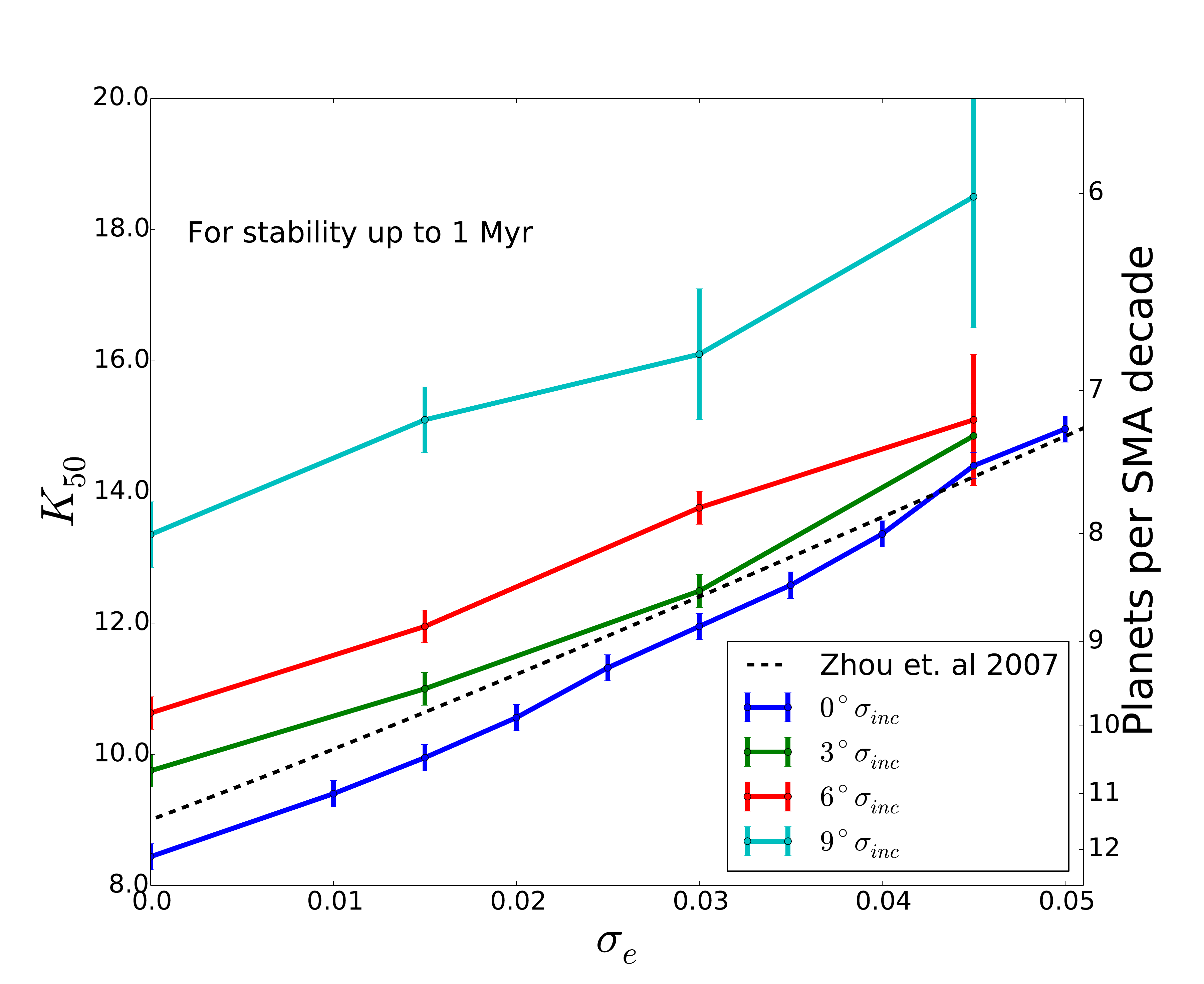}
\caption{The critical spacing $K_{\rm 50}$ as a function of orbital eccentricity, at time  $1$ Myrs
($\tau = 10^{7.5}$).  Different curves stand for  different mutual inclinations. For comparison,
we overplot in {\w dots} the result of \citet{Nanda} for coplanar systems.
}
\label{fig:figure5}
\end{figure}

Given the extra dimension in orbital parameters, it is only feasible to integrate these systems up to a time of one million years ($\tau = 10^{7.5}$). The survival probability for different types of orbits are shown in Fig. \ref{fig:figure5}. 

Individually, mutual inclination and initial eccentricity each has a strong destabilizing effect on stability. For the range of parameters that we explore, the critical spacing can be 
characterized as
\begin{equation}
K_{50} (\sigma_e,\sigma_{\rm inc}) \approx K_{50}(0,0) +
\left({{\sigma_e}\over{0.01}}\right) + \left({{\sigma_{\rm inc}}\over{0.04}}\right) \, ,
\label{eq:K50e}
\end{equation}
where $K_{50}(0,0)$ is the corresponding value for circular, coplanar systems for the same $\tau$ (eq. \ref{eq:K50}). An eccentricity dispersion of a mere one percent requires the spacing to expand by one Hill sphere, {\w for $\mu \sim 10^{-5}$}.  In relative terms, mutual inclination is less destructive to stability, by about a factor of $4$. 

One can also scale the above result in a natural unit of the problem, the Hill eccentricity,  $e_{\rm H} \equiv R_H/a$. For our parameters, ($\mu_1 \sim \mu_2 \sim 2\times 10^{-5}$), $e_{\rm H} \sim 0.024$. The expression for critical spacing, re-cast in these units, is,
\begin{equation}
K_{50} (\sigma_e,\sigma_{\rm inc}) \approx K_{50}(0,0) +
0.4 \times \left({{\sigma_e}\over{e_{\rm H}}}\right) + \left({{\sigma_{\rm inc}}\over{4.0 e_{\rm H}}}\right) \, .
\label{eq:K50eh}
\end{equation}
This leads to a surprisingly geometrical interpretation for our result. For every unit of eccentricity rise (in unit of $e_{\rm H}$), the average spacing between two adjacent planets ($K$) have to rise by $2.5 R_H$ in order to maintain stability; however, such a rise in eccentricity also reduces the closest approach (measured from the apo-apsis of the inner planet to the periapsis of the outer one) by $\sim 2 e_{\rm H} a = 2 R_{\rm H}$. So a system of mildly eccentric planets behaves similarly as a circular system that have the same closest approach. {\w We do not have a ready explanation for this result.}

In the following, we assume that the survival stability obtained at these shorter integrations can be extrapolated to longer times, in the form of Equation \refnew{eq:K50}.  

\subsection{Discussion: Extrapolating to 1 Gyrs?}
\label{subsec:extrapolate}

We do not yet understand the origin for the scaling in equation \refnew{eq:K50}. So it may be prudent to ask: can the scaling be extrapolated to as long as one billion years? For instance, in two-planet systems, it is known that beyond certain spacing the system can remain eternally stable. If high-N planetary system exhibits the same behaviour, equation \refnew{eq:K50} will break down at a certain spacing.


Lacking analytical understandings, we turn to numerical simulations for insight. \citet{SmithLissauer} experimented with equal-mass, constant-spacing 3-planet (or 5-planet) systems and found that as planet spacing reaches $K\sim 7$ ($\sim 8$ for 5-planet), the stability time, clocked at $10^7$ dynamical time at this point, starts to lift off and increases with $K$ much more steeply than Eq. \refnew{eq:K50}. It appears that this equation is violated beyond $10^7$ dynamical times, and indefinite stability at large spacing appears possible.

However, our results here suggest the opposite. We have integrated non-equal-mass, non-uniform spacing 7-planet systems for up to $10^{9.5}$ dynamical times, while still observing that Equation \refnew{eq:K50} holds well.  There is no unusual transition that occurs at $10^7$ dynamical times.  We suspect that our heterogeneous set-up (a dispersion of mass and spacing) opens up a more complex phase-space. Such dynamical system may not exhibit a single characteristic timescale, as the more restrictive cases in \citet{SmithLissauer}. In which case, it may be reasonable to extrapolate Eq. \refnew{eq:K50} by one order of magnitude, to 1 Gyrs ($10^{10.5}$ dynamical time).





\subsection{Discussion: Effects of Mean Motion Resonance}
\label{subsec:MMR}

\begin{figure} \centering \includegraphics[width=0.45\textwidth,trim=0 100 0 100,clip]{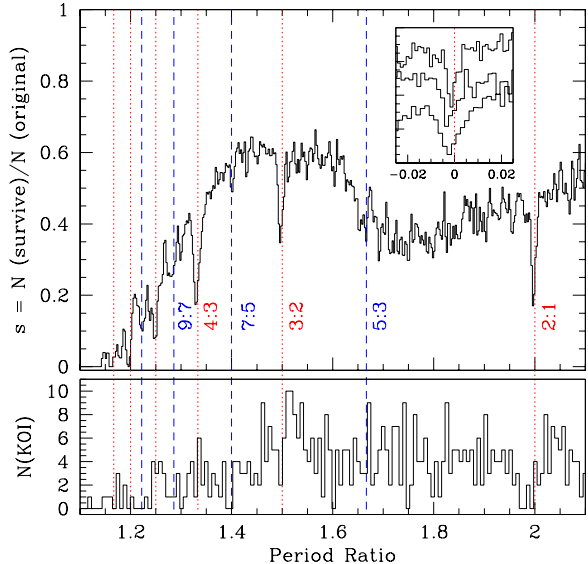} 
  \caption{The upper panel depicts the survival rate vs. period ratios for all pairs of planets in circular, coplanar simulations. Here, the survival rate is the ratio of stable pairs (at $10^7$ yrs) 
    The locations of the first order (dashed red lines) and the second order resonances (dashed blue lines) coincide with reduced survival rates, showing that mean-motion resonances tend to de-stabilize systems. The little inset at the top-right focuses on regions of width $\pm 2.5\%$ around 3 first order MMRs (2:1, 3:2, 4:3). There is a definite asymmetry: systems on the near side of resonances are preferentially depleted relative to those on the far side.
    The lower panel is the observed period ratios among all Kepler systems.  These also show a relative deficit of systems near these resonances in the near-side. The similarity between the two is intriguing.} \label{fig:MMR} \end{figure}

Previous studies \citep{Chambers,Funk} have identified that mean motion resonances tend to be disruptive for dynamical stability. This is also demonstrated in more detail in \citet{Nikhil} where they showed that, when the positions for Kepler-11 planets are shifted away from their current values, they may encounter mean motion resonances and become destabilized.  Here, using the 8000 circular, coplanar systems that we have produced, we perform a statistical investigation on the effect of resonances on stability (Fig. \ref{fig:MMR}).

It appears that, by $10^7$ yrs, systems that contain near resonant pairs are preferentially destroyed, compared to neighbouring systems. This is an order unity effect for first-order resonances, and weaker, albeit still visible, for second order resonances. Similar carving is also seen in our eccentric, inclined ensembles.

Amongst Kepler’s multi-planet systems, there has been an observed tendency for planets to ‘pile up’ just wide of mean motion resonances \citep{2011Natur.470...53L}. A number of mechanisms have been proposed for this asymmetry, including resonant repulsion \citep{2012ApJ...756L..11L,Batygin,Delisle}, stochastic migration \citep{2012MNRAS.427L..21R}, adiabatic mass accretion \citep{Petrovich}, and {\w planetesimal interactions \citep{Chatterjee15}}. We observe that (inset in Fig. \ref{fig:MMR}), for circular, coplanar systems, systems with period ratios just inward of MMRs are more severely carved relative to those on the far side of resonances. The asymmetry is more pronounced in slightly eccentric, inclined systems.
This asymmetric carving may present another explanation for the observed systems.
More investigation is planned in the future.

\section{Orbital Spacing of Kepler’s Multi-Planet Systems}
\label{sec:2}

\subsection{Sample Definition}

Observational data from the \Kepler Space Telescope was retrieved from the NASA Exoplanet Archive in June 19, 2014. All planet candidates with a KOI disposition of "False Positive" were removed. We are only concerned with high-multiple systems ($N_p \ge 4$ planets)
and our sample include 4 six-planet systems, 18 five-planet systems and 55 four-planet systems. 
There are a total of $257$ adjacent pairs, with {\w $92$} pairs in $5P$ and $6P$ systems alone.

\subsection{The intrinsic $K$ distribution}
\label{subsec:intrinsic}

To compare against theoretical results in \S \ref{sec:tight}, we require the intrinsic $K$-distribution (as opposed to the apparent one) of Kepler's multi-systems. However, such a distribution is not straightforward to obtain, due to three factors. First, the $K$ parameter depends on planetary masses, which are not known for vast majority of planets. Secondly, the presence of non-transiting (‘missing’) planets, either because they are too small to be detectable, or because they are unfortunately inclined relative to our line-of-sight, will cause the observed $K$ to be artificially inflated. Thirdly, {\w one must account for }selection effects. The KOI sample is severely incomplete for planets with period outward of $\sim 200$ days \citep{2013PNAS..11019273P,Silburt}. Systems that have tighter-than-average spacing will be preferentially picked up as high multiple systems, while those with larger spacings will not. This tends to skew the apparent $K$-distribution toward smaller $K$ values in these systems.

\begin{figure} \includegraphics[width=0.495\textwidth,trim=0 0 0 0,clip]{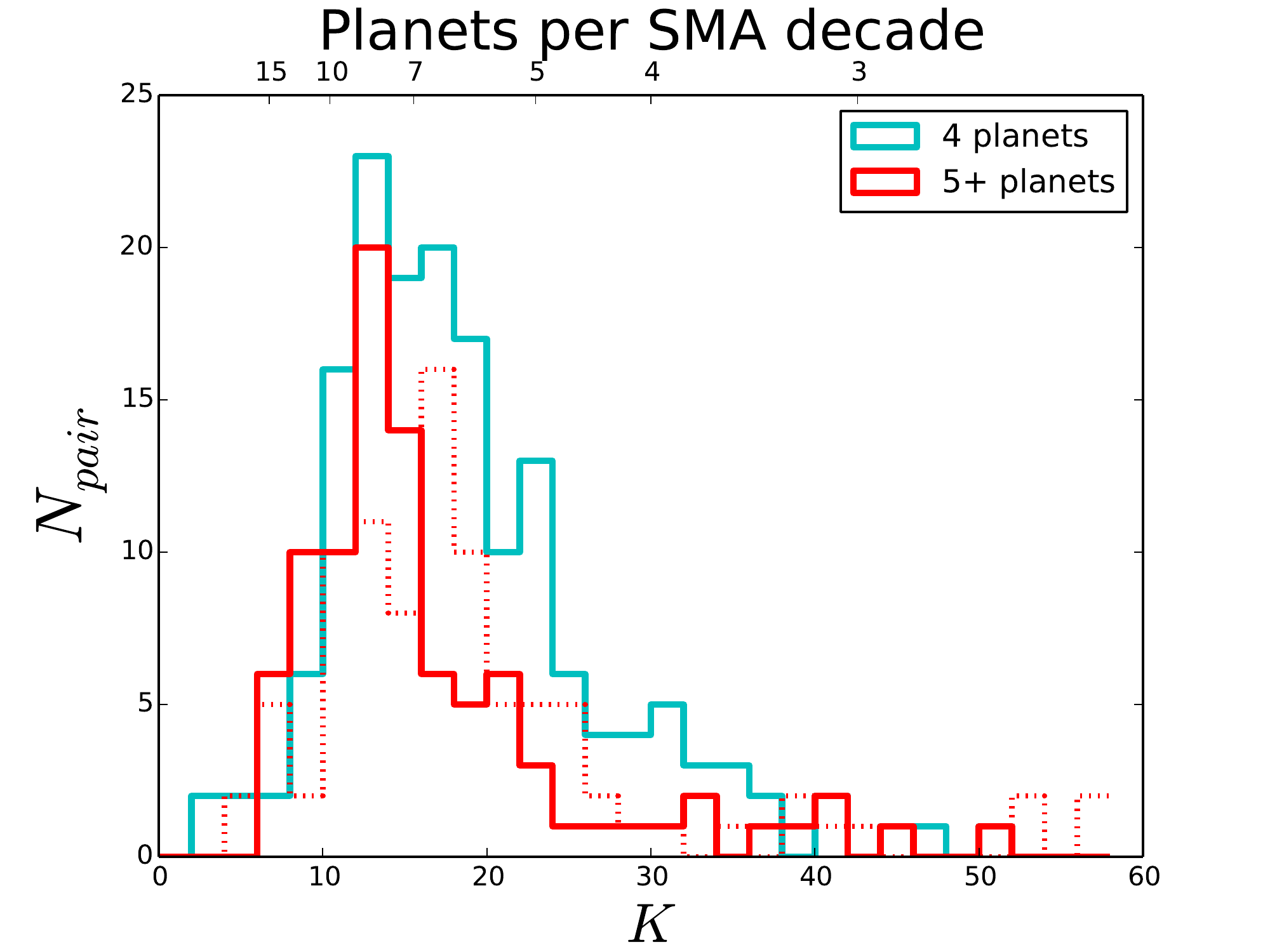} \caption{The observed $K$-distribution of high-multiple Kepler systems are shown as solid histograms for the 4P and 5P+6P systems. The mass-radius relation is assumed to be  $M \approx 3 M_\oplus (R/R_\oplus)$. According to KS test, the two histograms are statistically different, with the 5P+ systems more tightly spaced. The dotted curve shows the $5P+$ systems when we adopt instead $M = M_\oplus (R/R_\oplus)^{2.06}$, a scaling obtained for Solar system planets \citep{Lissauer2011b}. This leads to a more diffuse $K$-distribution.
}
\label{fig:figure1}
\end{figure}

We overcome the first hurdle by adopting the mass-radius relationship as obtained for planets that have measured masses.  Recent studies using both transit timing variation \citep{2012ApJ...761..122L,HaddenLithwick} and radial velocity \citep{Weiss} techniques have shown that planet masses scales approximately linearly with planet radii, or $M/M_\oplus \approx 3.0 R/R_{\oplus}$. {\w This applies to planets with sizes of a few Earth radii, the vast majority of our sample, and recent studies show that this relation may even extend to planets as small as $1.4 R_\oplus$ \citep{Dressing}. }
This {\w relation} differs from the one that applies to Solar system planets \citet{Lissauer2011b}. We plot the resulting apparent $K$-distribution in Fig. \ref{fig:figure1}.  There appears to be a sharp rise in the number of pairs above $K = 10$, and a sharp drop beyond $K \sim 22$, with some $75\%$ of pairs spaced between $10 \ge K \ge 22$.
In addition, the apparent $K$-distribution for $4P$ systems and that of $5P, 6P$ systems are statistically different (KS test returns a value of $0.003$),
with the former peaking at $K \sim 17$ while the latter around $K\sim 13$. It appears that even 4P systems may be heterogeneous, containing both tight spacing systems and large spacing systems. 
This prompts us to focus our attention exclusively to $5P$ and $6P$ systems in subsequent sections. 
Lastly, we notice that our adopted mass-radius relation leads to a sharper peak in $K$-distribution, when compared to other power laws.


To overcome the second and the third hurdles, it is necessary to ``observe'' synthetic populations to extract their apparent $K$-distributions.  We now describe our procedure in constructing and 'observing' these synthetic populations. 

Each planetary system orbit around a central star that is Sun-like, and have planets whose radii are distributed based on the determination in \citet{2013PNAS..11019273P,Silburt}, namely, with equal probability per logarithmic bin between $1$ and $2.5 R_\oplus$, and a linearly decreasing probability between $2.5$ to $15 R_\oplus$.  The mass of each planet is then determined using our default mass-radius expression.  {\w The orbital inclination, measured relative to a fixed plane, satisfies} a Rayleigh distribution with scale parameter $\sigma_{\rm inc}$ (eq. \ref{eq:probe}). Planets are placed in circular orbits, and all Euler angles are uniformly distributed.
The period of the inner-most planet in each system is drawn from a Rayleigh distribution with a scale parameter of 5 days. {\w This distribution is a good representation for our sample of planetary systems, but our conclusion is not sensitive to this particular choice.} To assign periods to the remaining planets, we assign a $K$ value that satisfies Equation \refnew{eq:probK} with a mean $K_{\rm mean}$ and a variance $\sigma_K$.  Each system is assumed to extend out to well beyond the observing boundary.

{\w With these choices, our average system, with $P_{\rm in} \sim 5$ days and $m \sim 5 M_E$, 
contain roughly 7 planets (ranging from 6-10 planets) inward of $200$ days.}
For the entire ensemble, the period distribution is flat in logarithmic space, consistent with observational determinations \citep{Silburt}.

We then identify planets that transit (impact parameter smaller than the stellar radius).  Not every planet that transits may be identified by the Kepler pipeline. We correct for this by adopting the detection completeness as a function of planet size and orbital period, as obtained in Fig. 2 of \citet{Silburt}. This effectively limit the detection sphere to periods inward of $\sim 200$ days, and miss some small planets.

Our procedure therefore accounts for both the missing planets and the selection bias.  The $K$-distribution from these 'detections' are then compared against that in Fig. \ref{fig:figure1}. We find that the best fit, using only 5P+ systems, is obtained when $K_{\rm mean} = 14, \sigma_K = 3.4$.  This corresponds to an effective spacing of $K_{\rm eff} = 12.3$. 

\begin{figure} \centerline{\includegraphics[width=0.49\textwidth,trim=0 0 0 0,clip]{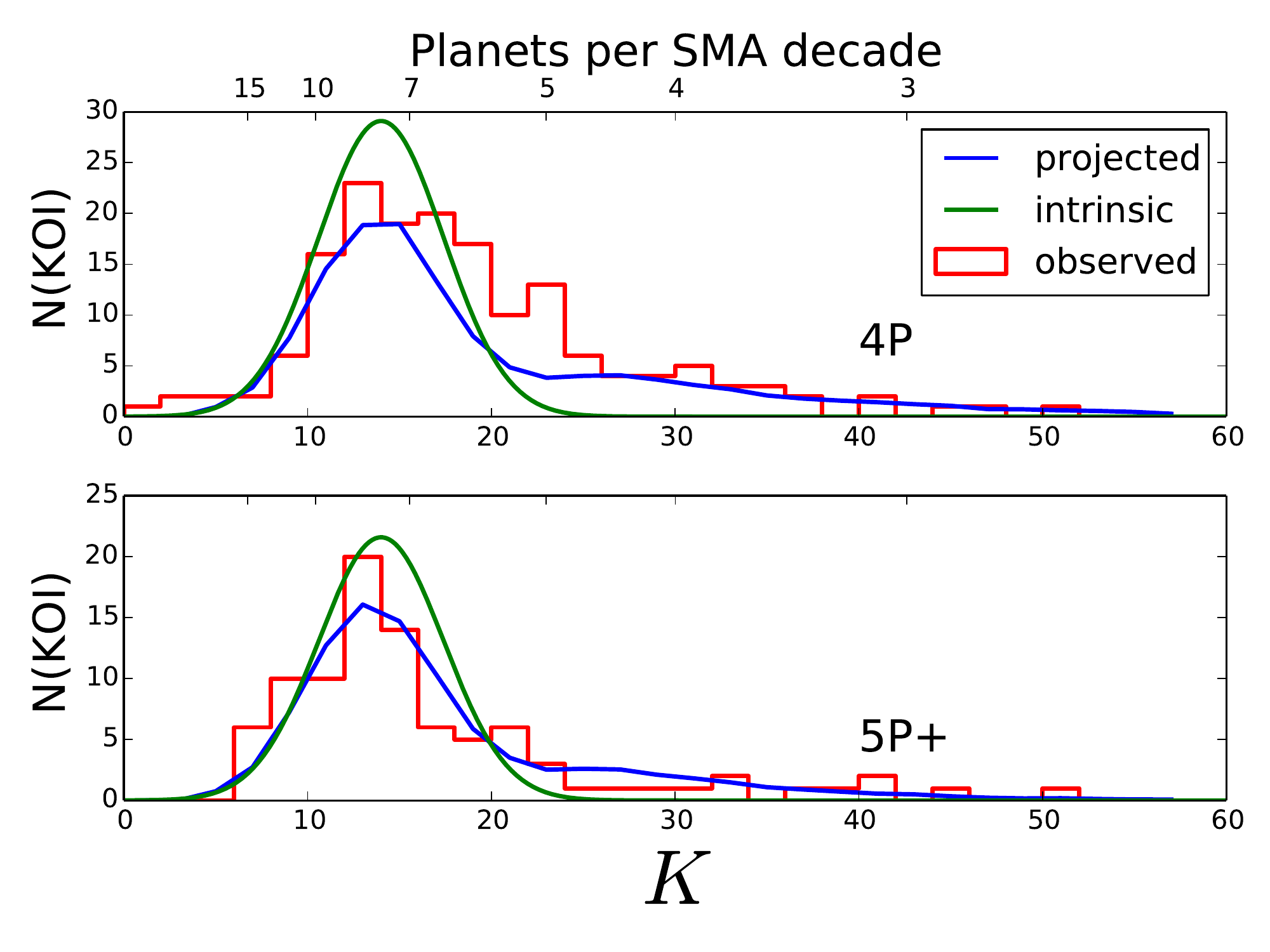}}
  \caption{The underlying spacing distribution. The observed spacing, for 4P and 5P+ systems respectively, are shown as solid histograms, the green curve is our best-fit solution for the underlying distribution of the 5P+ systems, and the blue curve its projection into observable space by synthetic observations. Our best fit, $K_{\rm mean}=14, \sigma_K=3.4$, for the 5P+ systems fail to account for some of the more widely spaced 4P pairs. 
Here,  we adopt $\sigma_{\rm inc} \sim 2\deg$ -- the $K$-distribution does not vary much when the inclination dispersion is varied. 
} \label{fig:3} \end{figure}

As is shown in Fig. \ref{fig:3}, the 'observed' $K$-distribution is always broader than its intrinsic counterpart. The broadening is caused by intervening planets missed by transit surveys:
our simulations show that for low inclinations ($\sigma_{inc} \sim 2 \deg$), $\sim 35 \%$ of the apparent 5P and 6P systems contain one or more missing planets hidden in-between the transiting planets.

Moreover, the apparent tight spacing in high multiples is not a result of selection bias, namely, only systems that have intrinsically the tightest distribution can fit inside the detection sphere ($\sim 200$ days) and be observed as a high-multiple system (see \S \ref{sec:discussion}).



We can translate the $K$-distribution to a more intuitive unit, the number of planets within a decade of semi-major axis, typically located from 0.1 to 1.0 AU for our case. This number is  marked as 'planets per SMA decade' in multiple figures. 
The \Kepler 5P+ systems contain between 6 to 10 planets per SMA decade, with a mean of $\sim 8$. 
Contrast this with only 4 planets per decade in the terrestrial region of the Solar system.


{\w Our determination of the intrinsic $K$-distribution is not affected by our choice of inclination dispersion, perhaps a bit paradoxically. A higher dispersion means more planets in a given system escape detection in a transit survey, and this should have increased the average $K$ value. However, this also disqualifies such a system from our 'high multiple' sample. As a result, the intrinsic $K$-distribution we obtain, by focussing on high multiples only, is not much affected.}




\section{Discussion}
\label{sec:discussion}

Gyro-chronology studies, the dating of stars using their stellar rotation periods, reveal that stars in the Kepler field, at least ones that have measurable rotation periods, have typical ages of a couple Gyrs \citep{Walkowicz,McQuillan}.  This motivates us to assume a default age of 1 Gyrs for all Kepler systems. This may be an underestimate in many cases. Older stars are much less magnetically active and therefore less variable. They also rotate with longer periods. As a result, they may not be selected for gyro-chronology studies and results from the latter may be biased towards younger stars. Fortunately, an older age simply strengthens the argument we present below.

If all Kepler high multi systems are circular, coplanar, equation \refnew{eq:K50} predicts that $50\%$ of the planetary systems can remain stable to beyond $t= 10^9$ yrs ($\tau\sim 10^{10.5}$) if $K_{\rm eff} \geq 10.2$. This is surprisingly close to our best fit for Kepler high multis, $K_{\rm eff} = 12.3$.

The proximity is even more striking if one allows for just a small amount of eccentricity and non-coplanarity in the systems.  Adopting the measurements by \citet{WuLithwick,HaddenLithwick} for TTV pairs, $\sigma_e \approx 0.02$, equation \refnew{eq:K50e} requires that $K_{\rm eff} \geq 12.2$.

This observation, that Kepler high multis are perched at the cliff of instability {\z \citep[see also][]{Lovis, FangMargotPacking}}, as we suggest below, is most likely the result of continuous sculpting. This suggestion is the main thesis of this work.

Our intrinsic distribution exhibits {\w a sharp} fall-off  at larger $K$ (see Fig. \ref{fig:3}), similar to the observed population. We believe this fall-off is genuine, not a result of observational bias.
When synthetically observing populations of planets with a range of $K_{\rm mean}$ and $\sigma_K$, we find that even systems with spacing as large as $K_{\rm mean} = 17.5$ (fixing $\sigma_K= 3.5$) can be observed as 5P+ only $40\%$ less frequently than our best fit ($K_{\rm mean} = 14$). Their presence should therefore be detectable.
The lack of such wide spacing system in the observed high multis is therefore likely genuine.


\subsection{Dynamical Scuplting}
\label{subsec:initialk}

The formation of \Kepler planets themselves is not yet understood and different theories have been proposed {\w \citep{TerquemPap,IdaLin6, 2012ApJ...751..158H,ChiangLaughlin, Chatterjee13}}. However, it looks likely that they were made in the first few Myrs. This is corroborated by the presence of hydrogen envelopes on these low-mass planets, for both Neptunes \citep[e.g.][]{Borucki} and super-Earths \citep{WuLithwick,Weiss}. 

How does such a formation time jibe with the orbital spacing of the observed systems?  The nascent disk can weed out (and possibly reform) systems that are unstable within a few million years ($K_{\rm eff} \leq 8.1$, eq. \ref{eq:K50}), but it should not discriminate among systems that have spacing wider than this.  On the other hand, while it is natural that only systems with spacing wide enough to be stable over their lifetimes can be observed, it is striking how the observed systems skirt the edge of stability. 

This chain of thought then leads us to propose the following scenario. Planetary systems were initially formed with a range of $K_{\rm eff} \geq 8.1$. This distribution is continuously sculpted by dynamical instability. So by 1 Gyrs, the remaining population will exhibit a sharp cut-off at the border of stability.  We call this 'the sculpting hypothesis'.





Under this hypothesis, it is possible to reconstruct the initial planet population, after we posit a number of assumptions, some more sound than others. We assume that the all systems have the same small $K$-dispersion as we observe in today's high multis, $\sigma_K = 3.4$. Since orbital spacing in a system likely reflects its formation environment, we expect the spacings to be similar within a given system.  We further assume that all systems at birth have the same dispersion in eccentricity as the currently observed TTV sample, $\sigma_e \sim 0.02$, and a comparable dispersion in inclination.  To 'reverse' the effect of carving, we augment the current planet population by three populations that are similar in number but shifted to a smaller $K_{\rm eff}$ by $0.7\times \log(t/{\rm 1 Gyrs})$, for the three logarithmic time  intervals between $t=10^6$ and $t=10^9$ yrs.  The value $0.7$, according to our numerical experiments, is the range of spacing that is carved away by the passing of each logarithmic interval. And the initial $t=10^6$ yrs is the putative lifetime of our nascent disks. 

\begin{figure} \centering \includegraphics[width=0.49\textwidth,trim=0 0 0 0,clip]{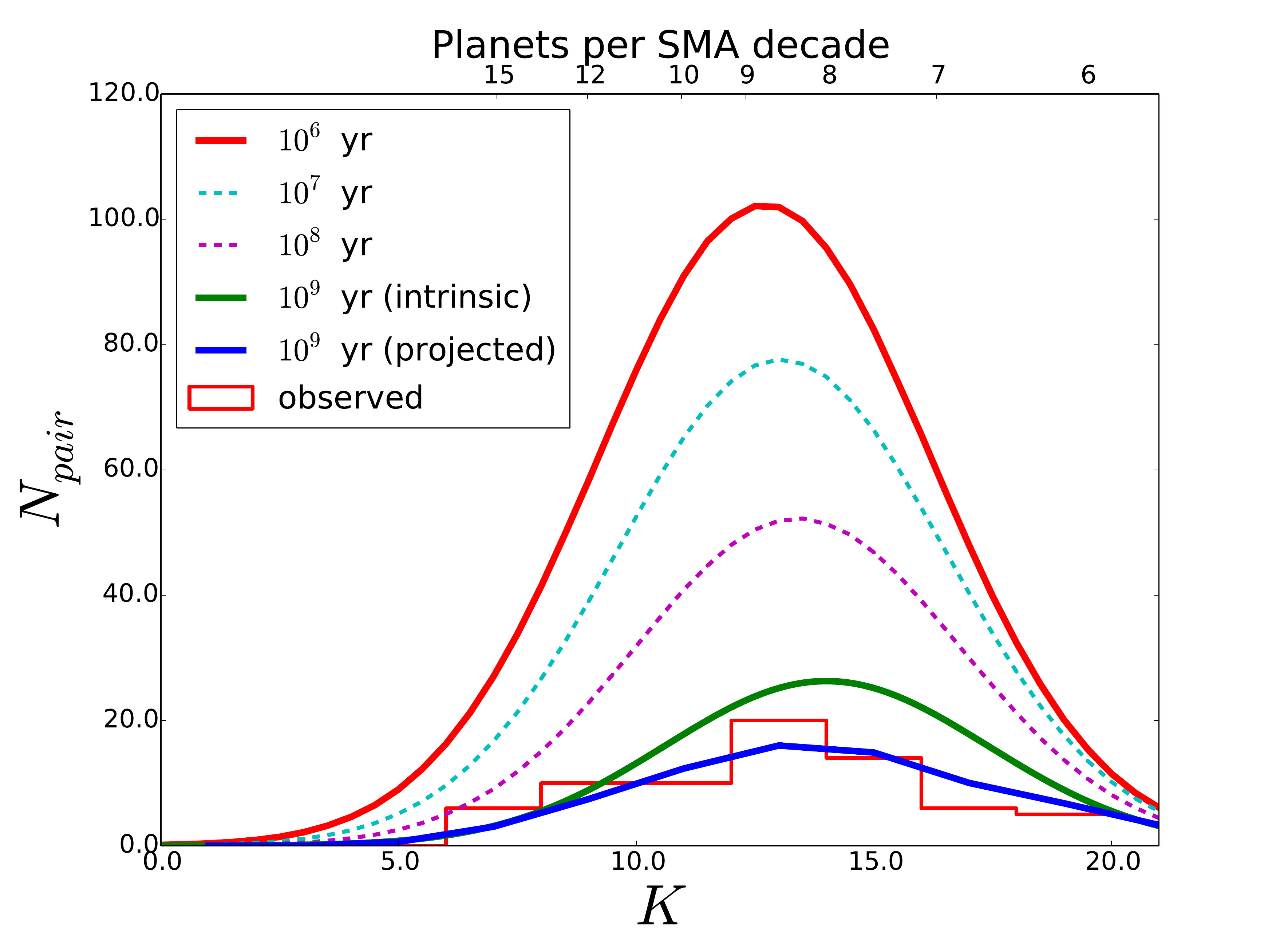} \caption{The sculpting hypothesis. The solid histogram and the solid curves are those displayed in the bottom panel of Fig. \ref{fig:3}, namely, the observed (histogram) and the underlying (green curve) spacing distributions for \Kepler 5p and 6p systems. The dashed curves, from bottom to top, represent our reconstruction of the earlier spacing distributions at, $10^8$, $10^7$ and $10^6$ years, respectively. See text for details. The primordial population (posited to be the one at $10^6$ yrs) are in average more tightly spaced, and contain roughly 4 times more systems than those remain today.
}
\label{fig:figure6}
\end{figure}

The resulting initial population is illustrated in Fig. \ref{fig:figure6}. In this model, by 1 Gyrs of age, roughly $75\%$ of the original systems have been disrupted. This fraction could rise or drop depending on what we assume for the initial $K_{\rm eff}$ distribution.

What happen to systems that experience dynamical instability, in the interval since their birth and today? Once two planets are excited to crossing orbits, they will physically collide with each other after a short time, as determined by the area of the planet subtended on the sky,
\begin{equation}
t_{\rm coll} \sim {{4\pi a^2}\over{\pi R_p^2}} \, T_1 \sim 10^6 {\rm yrs} \, \left({a\over{0.1 {\rm AU}}}\right)^2 \left({{R_p}\over{R_\oplus}}\right)^{-2} \left({{T_1}\over{12 {\rm days}}}\right)\, .
\label{eq:tcollision}
\end{equation}
{\w This expression assumes that the planets have acquired significant eccentricities and inclinations {\w before }impact, and that there is little gravitational focussing. Both assumptions are born out by our numerical simulations. In particular, the latter assumption is reasonble because at} collision, their relative velocities can be higher than their surface escape velocities \citep[$\sim 10-20 \km/\s$, see, e.g.][]{WuLithwick}, since the Keplerian orbital speed is of order $100\km/\s$ at the relevant distance. Such a collision will likely be destructive, as opposed to conglomerative \citep{Asphaug,Leinhardt}. Some planets may be lost and massive debris may be produced. The former will increase the spacing within the system, while the latter will tend to cool down the orbit and prevent further instabilities from happening. At the moment, we are incapable of predicting the outcome of such a collision, but speculate that this may explain the abundant low-multiple systems one observes in the \Kepler field.

\subsection{Formation Environment}
\label{subsec:formation}


We speculate briefly on the physical environment that can give rise to the compact spacing of \Kepler systems. If our sculpting hypothesis proves correct, this is not restricted to only the high multiple systems, but likely applies to all close-in planetary systems. 

The orbital spacings between terrestrial planets are wide, with $K \sim 30-40$.\footnote{However, even this wide spacing, given the presence of giant planets and orbital eccentricities, still puts the Solar system at the edge of instability.} These values have been reproduced by a number of simulations where these planets conglomerate out of a sea of smaller bodies \citep[e.g.,][]{Chambers01,Kokubo06,OBrien06,Morishima10}. In these simulations where there is zero or little dissipation acting on the planetary orbits, the proto-planets can stir each other up to of order their surface escape velocities. Their orbits, being dynamically hot, require larger spacing to avoid further disruption. This reasoning is supported by simulations of \citet{Kominami02}, which we discuss below.


Starting from a population of Mars-sized proto-planets, but exerting an artificial dissipation on the planet eccentricity with timescale $\tau_{\rm damp}$, \citet{Kominami02} showed that, terrestrial-like systems ($K \sim 30$) can be produced if $\tau_{\rm damp}$ is relatively weak (more than a million times the orbital period), and closely-packed {\w systems result} if damping is strong. When $\tau_{\rm damp} \sim 10^5$ orbital time, in particular, the simulations produce systems that are spaced between $8$ to $13 R_{\rm H}$ -- much like the initial population that we inferred in Fig. \ref{fig:figure6}.  At $0.1$ AU, this corresponds to a damping time of $\tau_{\rm damp} \sim 3000$ yrs.  In \Kepler systems, the primordial disk is likely more massive than the terrestrial system studied by \citet{Kominami02}, so the requisite e-damping time may be even shorter. To put this timescale into context, the expected e-damping time from Type-I migration in a minimum mass solar nebula is $\tau_{\rm damp} \sim 3$ yrs \citep{Tanaka04}.


So if our sculpting hypothesis is correct, we infer that the formation environment of the \Kepler planets have to be {\w moderately} dissipative, {\w with an eccentricity damping timescale that is longer than the Type-I damping timescale, but much shorter than the} million-year lifetime of the nascent disks.

\subsection{Constraints on Orbital Shapes}
\label{subsec:einc}

We can look at our results in a different way. The observed spacing could be used to constrain the values of 
eccentricity and mutual inclination in (and only in) these high multiple systems. Inverting equation \refnew{eq:K50e}, adopting $K_{\rm 50}(0,0) = 10.2$ (required at 1 Gyrs) and $K_{\rm 50}(\sigma_e, \sigma_{\rm inc}) = 12.3$ (observed at 1 Gyrs), we find the following independent limits,
 \begin{equation} 
\sigma_e \leq 0.02\,,\,\,\,\, \,\,\,\,  {\rm and},\,\,\,\,\,\,\,
\sigma_{\rm inc} \leq 0.084\, \,  {\rm radian}\, \approx 5 \deg\, .
\label{eq:einc}
\end{equation}
{\w Although this constraint is obtained for our choice of mass-radius relationship, it likely applies even when the latter deviates from our choice.}
Similar constraints on orbital shapes were also obtained by \citep{Nikhil} for a specific planetary system, Kepler-11. If $\sigma_{\rm inc} \sim \sigma_e/2$, as characteristic of 2-body relaxation in disks \citep{IKM}, we would obtain $\sigma_e \leq 0.02$ and $\sigma_{\rm inc} \leq 0.01 \sim 0.6\deg$.  These are indeed very cold dynamical systems.

How do these constraints compare to what we have found via other channels? Most previous studies have not isolated high-multiple systems from the general catalogue and their results are therefore not directly comparable, though a number of studies of planet mutual inclinations 
have indicated low values {\w \citep[e.g.][]{FangMargotInclination,TremaineDong,Fabrycky14}}.

There are only a few studies on planet eccentricities. Studying the distribution of transit durations in Kepler candidates have led \citet{Moorhead} to conclude that $\sigma_e \sim 0.2$. However, this result is heavily plagued by uncertainties in stellar radius determinations and is likely an over-estimate. TTV studies \citep{WuLithwick,HaddenLithwick}, on the other hand, have led to a much smaller eccentricity dispersion ($\sigma_e \sim 0.02$), for a sub-population of planet pairs that are near mean-motion resonances. In fact, this small but non-zero dispersion partially motivates us in claiming that the observed systems lie on the edge of instability.  TTV studies also reported the existence of another sub-population where the eccentricities are substantially higher ($\sigma_e \sim 0.1$), but these likely occur exclusively in low-multiple systems.


\section{Conclusion}

The architecture of Kepler multiple systems contains important pieces of clue for planet formation. In this work, we show that high-multiple Kepler systems have spacings that narrowly afford them dynamical stability in the giga-year timescale. This leads us to suggest that many more planetary systems were initially formed with even smaller spacing, but they have been since whittled down by dynamical instability, possibly producing some of the low-multiple Kepler systems we see today.
If this is true, the final formation stage of these planets proceed in highly dissipative environment, with eccentricity damping timescales much shorter than the disk lifetimes.

Along the way, our numerical explorations have uncovered a number of interesting features in the stability of high-N systems.  Consistent with previous investigations, we find that system stability time, quantified here as the time by which $50\%$ of the initial systems remain stable, rises {\w sharply} with the spacing. When the spacings in a given system are not uniform but exhibit a spread, one can easily quantify the effect by adopting an effective spacing, $K_{\rm eff} = K_{\rm mean} - 0.5\sigma_K$. We give fitting formula for stability time when the planetary orbits are slightly eccentric and/or mutually inclined. Eccentricity, even at the level $1\%$, affects critical spacing significantly.  As a result, we are able to place a stringent limit on the eccentricity of high-multiple Kepler systems, $\sigma_e \leq 0.02$. We find that systems that contain mean-motion-resonance pairs are preferentially destabilized. 

A large number of both theoretical and observational questions remain unanswered. 
What provides the strong eccentricity dissipation in planet formation? What is the origin of dynamical instability in high-multiple systems? What are the final outcomes in these dynamically unstable systems? How do these outcomes compare against the observed systems?

\section{Acknowledgements}
{\w We thank the anonymous referee for a meticulous and thoughful set of comments.}
BP thanks NSERC for an undergraduate summer fellowship, and Nikhil Mahajan for assistances. YW acknowledges the NSERC funding, as well as conversations with Yoram Lithwick, Doug Lin (who pointed out that Johannes Kepler did ponder about the planet packing problem) and Scott Tremaine.

\bibliographystyle{apj}


\end{CJK*}

\end{document}